\documentclass[16pt,preprint]{aastex}

\begin{document}
\title{A ``pulsational'' distance determination for the Large Magellanic 
Cloud\altaffilmark{1}
}
\author{Marcella Marconi}
\affil{INAF - Osservatorio
Astronomico di Capodimonte, via Moiarello 16, I-80131 Napoli,
Italy. email: marcella@na.astro.it}
\author{ Gisella Clementini} 
\affil{INAF - Osservatorio Astronomico di Bologna, Via 
Ranzani 1, I-40127 Bologna, Italy. email: gisella.clementini@bo.astro.it}
\altaffiltext{1}{Based on observations collected at the European Southern
Observatory, proposals 62.N-0802, 66.A-0485, and 68.D-0466}

\begin{abstract}

We present results from the theoretical modelling of the observed $B,V$ 
light curves of 14 RR Lyrae stars in the Large Magellanic 
Cloud. The sample includes 7 fundamental and 7 first overtone pulsators covering the
metallicity range from $-$2.12 to $-$0.79 dex in [Fe/H], with average value of $-$1.54 dex.
Masses, intrinsic luminosities, effective temperatures, and reddenings were 
derived by fitting high accuracy multiband 
light curves available for these RR Lyrae stars  to
nonlinear convective
pulsation models. 
Individual distance moduli were determined for each variable, they lead to
an average distance modulus for the Large Magellanic Cloud 
of  $\mu_{0}$= 18.54 $\pm$ 0.02 ($\sigma$= 0.09 mag,  
standard deviation of the average), 
in good agreement with the 
``long" astronomical distance scale.

\end{abstract}

\keywords{Stars: evolution - stars: horizontal branch - stars:
variables: RR Lyrae - LMC}

\section{INTRODUCTION}

One of the most debated issue of  modern astrophysics is the
definition of the local distance scale and of the distance of its first step, 
 the Large
 Magellanic Cloud (LMC), in particular. Indeed, current calibrations of 
the extragalactic distance scale and consequent evaluations of the 
Hubble constant rely upon the Classical Cepheid PL relations with
 the zero point fixed by some assumption on the LMC distance (see e.g.
Freedman et al. 2001, Tammann, Sandage, \& Reindl 2003).
Recent determinations of the LMC distance modulus, either based on
 Population I or II distance indicators, cover the range 18.3-18.7
 mag (see discussions in Cacciari 1999, Walker 1999, Carretta et al. 2000, 
 Caputo et al. 2000). However, in the last couple of years different distance 
 indicators have followed the trend to converge
to a common value of $\mu_{LMC}$=18.52 $\pm$ 0.02 mag, on the basis of updated 
accurate photometric data for both RR Lyrae and red clump stars, as well as 
revised techniques and consistent reddening evaluations (see discussions in 
Clementini et al. 2003, hereinafter C03, Walker 2003, and Alves 2004). 

Among the different methods exploiting the radially pulsating stars as standard
candles a quite
innovative  and promising technique is represented by the fitting of
the observed light and radial velocity curves of a pulsating star 
to those predicted by nonlinear
pulsation models  (see e.g. Wood, Arnold \& Sebo 1997, hereinafter WAS; 
Keller \& Wood 2002; Bono, Castellani, \& Marconi 2000, 2002, hereinafter
BCM00 and BCM02, respectively; Di Fabrizio et al. 2002).  This technique is based on the
direct comparison between observed and  predicted  curves,
rather than involving related parameters such as pulsation amplitudes
and Fourier parameters, and performs the theoretical
reproduction of both general features (period, shape, 
amplitude) and morphological details  such as bumps and humps, of
the observed curves. Given the sensitivity of the curve
detailed morphology to the model input parameters, as well as to the physical and numerical assumptions, the fitting of
actual light and radial velocity curves offers a unique opportunity  to
obtain sound estimates of the intrinsic stellar properties of
the variables, providing at the same time a key test of the
theoretical predictions.

High accuracy observational data,
consisting of well sampled multiband light and, possibly, 
radial velocity curves, knowledge of the related 
pulsation characteristics, namely period, colors and amplitudes, and of the 
metallicity 
of the variable star, 
are needed for a meaningful application of the method.
The output of the procedure is a set of models for fixed period and 
metal abundance and with varying 
the mass, the luminosity and the  effective temperature\footnote{We note 
that mass, luminosity and effective temperature are not independent variables 
because of the existence of a Period-Density relation for pulsating stars.}, among which the 
best fitting model is that best
reproducing the observed characteristics of the pulsation, within
acceptable values of the derived intrinsic parameters.

In the case of variables of RR Lyrae type this technique has already 
provided very good results for Galactic field first overtone RR Lyrae stars 
(see BCM00,
 Di Fabrizio et al. 2002),  whereas for the fundamental mode pulsators
 satisfactory results were obtained except for the long period
 variables close to the red edge of the RR Lyrae instability strip 
 (see Castellani, Degl'Innocenti, \& Marconi 2002, Di Criscienzo, Marconi, \&
 Caputo 2004). 
 
As for the Classical Cepheids a similar technique was early applied by 
WAS to the LMC bump Cepheid HV905.   More recently,  
relying on the same theoretical pulsational scenario as in BCM00, the method
was applied to two
LMC  fundamental mode Cepheids showing well-defined bumps  along the
decreasing (short-period) and the rising (long-period) branch  of the
light curve,  respectively (BCM02), based on
data from the OGLE catalogue  (available at
sirius.astrouw.edu.pol/$\sim$ogle/ogle2/dia).  A good fit was
obtained for both variables and for an LMC distance modulus ranging
from 18.48 to 18.58 mag (see BCM02 for details), in agreement with
results by C03. An independent application to  the MACHO 
$V$ and $R$ light curves of 20 {\it bump} Cepheids  in the LMC 
(including HV905) was 
performed by Keller \& Wood (2002), on the basis of the WAS nonlinear 
pulsation code. 
They obtained a mean LMC distance modulus of 
18.55 $\pm$ 0.02 (intrinsic error of the average).

Both in the case of RR Lyrae stars and Classical Cepheids the best fit
 models based on BCM00 theoretical scenario seem to suggest that the standard 
value of the mixing length 
parameter ($l/H_p = 1.5$) adopted to close the dynamical-convective 
equation system is suitable to reproduce the light variations of 
variables in the blue region of the instability strip. A higher value 
of $l/H_p$ is needed instead toward the red region of 
the fundamental mode instability strip (see e.g. BCM02). This occurrence is 
confirmed
by results based on other independent ``pulsational'' 
methods to constrain the RR Lyrae and globular cluster distance scale,  
in particular the Period-Amplitude relations and the comparison 
in the Period-Magnitude diagram 
(see Di Criscienzo et al. 2004).

In this paper we present results from the theoretical modelling of  the
observed $B,V$ light curves of 14 RR Lyrae stars  in the LMC, 
 based on the detailed
photometric study of the LMC RR Lyrae
stars by  C03 and Di Fabrizio et al. (2004, hereinafter DF04). Our sample includes 
7 fundamental mode (RRab) and 7 first overtone (RRc) pulsators that,  
according to Gratton et al. (2004, hereinafter G04) spectroscopic analysis, cover the
metallicity range from $-$2.12 to $-$0.79 in [Fe/H], with average value
of $-$1.54 dex.
The application of the method to a significant number of Population II
pulsators in the same stellar system, namely
the LMC RR Lyrae stars, represents a powerful test of the inner consistency 
and predictive capability of the adopted theoretical scenario, providing at the same
time an independent estimate of the LMC distance modulus, and
a direct comparison between  Population I and II distance indicators, 
which contributes to strengthen our knowledge of the local distance scale.

The organization of the paper is the following. In Sect. 2 we briefly
discuss  the adopted photometric (light curves and reddening values)
and  spectroscopic (metallicities) databases.  In Sect. 3 we present
the pulsation models, whereas the fit of the observed  light
curves and the derivation of the intrinsic stellar parameters (masses,
 luminosities, and effective temperatures) both for fundamental and first overtone
pulsators is described in Sect. 4. In this section we also discuss the 
comparison with the intrinsic parameters 
predicted for the program stars by the evolutionary  horizontal branch (HB) models, and by the
 Fourier decomposition of the visual light curves of the fundamental mode 
 RR Lyrae stars.  
 The derived  LMC distance modulus and the associated
uncertainty are presented in Sect.  5 along with a discussion of the
resulting RR Lyrae luminosity-metallicity relation.  A summary of the results and  some final remarks in
Section 6 close the paper.

\section{THE OBSERVATIONAL DATABASE}
The RR Lyrae stars analyzed in the present paper were selected from the
high quality photometric catalogue of variable stars in the LMC published by DF04. 
The catalogue contains 
$B,V,I$ light curves in the Johnson-Cousins photometric system of 
162 variables located in two areas 
(referred to as Field A and B)
close to the bar of the LMC,
among which 135 are RR Lyrae stars.
These variables were used by C03 to very carefully determine the distance
of the LMC, getting the value $\mu_{0}(LMC)$=18.515 $\pm$ 0.085 mag
(1 $\sigma$ scatter of the mean) as the average of 17 independent methods
based both on Population I and II 
distance indicators.
A large fraction of the RR Lyrae stars in DF04 were also analyzed spectroscopically by 
G04 who measured individual metal abundances for about
a hundred of them, using low resolution spectra taken with FORS at the
VLT. 

The target stars were selected by  visual inspection of DF04 catalogue, 
preferring objects with well sampled and accurate light curves, and variety 
of morphologies, periodicities, and colors.
We chose for our analysis a number of single-mode fundamental and first overtone pulsators
not affected by Blazhko effect (Blazhko 1907) or other irregularities of the
light variations. The selected objects all have evenly covered $B,V$ light curves with 
about 60-70 data points in $V$ and about 40 in $B$, and 
with internal photometric accuracy of 0.02-0.04 mag in $V$ and 0.03-0.05 mag in $B$, 
according to DF04. The $I$ light curves in DF04 catalogue 
are generally of lower accuracy and have a rather limited number 
of data-points (14-15), thus we did not attempt the fit in this band.

To ensure variety of the light curve morphologies we selected 
objects spanning a range in period: from 0.487 to 0.656 days
for the fundamental mode, and 
from 0.232 to 0.385 days for the first overtone pulsators.

Metallicities [Fe/H] are available for all our targets, either from the 
spectroscopic analysis by G04, or estimated according to 
Jurcsik and Kov\`acs (1996) method 
from the $\phi_{31}$ parameter of the 
Fourier decomposition of the
visual light curves (see Sections 6 and 6.1 of DF04). 
G04 spectroscopic metal abundances are tied to the metallicity scale defined
by Harris (1996) [Fe/H] abundance for the galactic globular clusters M~68 and 
NGC~1851 (see Section 4 in G04), on average they are 0.06 more metal-rich than    
the Zinn \& West (1984) scale. The ``photometric" metallicities 
(star \#9660, only) were transformed to G04 scale
according to the
procedure described in Section 3.3 of C03. 
The metallicity range spanned by the selected targets (from $-$2.12 to 
$-$0.79 dex in [Fe/H] in G04 scale) is large 
enough to allow the check of the luminosity-metallicity relation
of the LMC RR Lyrae derived by other studies (e.g. C03 and G04)  
using the absolute magnitudes derived from the model fitting (see Section 5.1).

Following the selection criteria 
described above, we ended up having in our sample a number of RRab's 
with relatively long period ($P > 0.6$ days). 
One of them (star \# 9660) also has relatively high metal abundance
([Fe/H]=$-$1.5) and, as confirmed by its rather red color ($<B-V>$=0.47 mag),
lies close to the red edge of the RR Lyrae instability
strip\footnote{According to C03 the fundamental red edge (FRE) of the
RR Lyrae stars in DF04 catalogue occurs at $<B-V> \sim $0.51 mag.}. 
This is the  object for which 
we have achieved the  least satisfying model fitting (see Section 4 and
Fig. 1).
Indeed, as anticipated in the Introduction, current pulsation models
often fail to reproduce the morphology of the light curves of 
long period fundamental mode RR Lyrae stars 
close to the FRE. 
In fact, while the theoretical light curves are generally able to 
reproduce   
 the saw-like shape with amplitude decreasing from the FBE 
to the FRE of the observed fundamental mode 
RR Lyrae stars, at the lowest effective temperatures the models also show a 
secondary peak before 
the phase of maximum light. This secondary peak is of increasing
 strength as the FRE is being approached. However, it is not present in the 
observed light curves.

C03 estimated the average reddening of the LMC fields containing
our targets using the colors of the edges of the 
 RR Lyrae instability strip. 
 They derived $E(B-V)$=0.116$\pm$0.017 
 for the field closer to the LMC bar (Field A), and  $E(B-V)$=0.086$\pm$0.017 mag
 for the more external one (Field B), corresponding  to a differential reddening of 0.03 mag
 between the two areas.  
 Our RRc sample includes two of the 3 variables that define the 
 first overtone blue edge (FOBE) of the RR Lyrae stars in C03 and DF04.
  C03 also estimated individual reddenings
 for the RRab variables  using the Sturch (1966) method, which is
 based on the mean $<B-V>$ colors at minimum light of the variables. 
The color excesses inferred 
from our model fitting procedure can be directly compared with both average 
and individual reddenings in C03, thus providing an independent check of C03
values, hence of the distance for  the LMC in that paper.
This check is particularly relevant in light of the controversy
existing on the actual reddening towards the LMC (see discussions in
Udalski et al. 1999, hereinafter U99; Zaritsky 1999; C03; 
Alcock et al. 2004, hereinafter A04; Zaritsky et al. 2004) and is presented in Section 4.4.   

The list of RR Lyrae stars analyzed in the paper is provided in 
Table~\ref{t:table1} along with a summary of their relevant observed
quantities.  Namely we list: identifier, variable type and period, 
intensity-averaged mean $<V>$ and $<B>$ magnitudes with the related
total uncertainties\footnote{The total photometric 
errors associated to the mean $<V>$ and $<B>$ magnitudes
are the sum in quadrature of the internal uncertainties,
corresponding to the 
average residuals from the 
Fourier best fitting models of the observed light curves as computed by 
GRATIS (GRaphical Analyzer of TIme Series, Di Fabrizio 1999, Clementini et 
al. 2000; see DF04), and of the contributions due to uncertainties of 
the photometric calibration 
(0.0175 and 0.032 mag in $V$ and $B$, respectively, C03 and DF04), and  
aperture corrections (0.018 and 0.019 mag in $V$, and 0.032 and 0.014 mag in 
$B$, in field A and B, respectively, C03).}, number of data 
points and amplitudes of the light curves, taken from DF04; metal abundances 
from G04; and reddenings taken from C03.

\section{PULSATION MODELS}

In the last few years an extensive and detailed set of updated
nonlinear convective pulsation models of RR Lyrae stars has been developed 
covering wide ranges in period and metal
abundances (see Bono et al. 2003, Marconi et al. 2003 and references
therein).  The physical and numerical assumptions adopted in the theoretical 
computations were widely discussed in these papers, to which the
interested reader is referred for details. 
We only remark here that thanks to the
nonlinearity and the inclusion of  convective transfer in a non local,
time-dependent approach (see Stellingwerf 1982, Bono \& Stellingwerf
1994 for details) current models are able to reproduce all the
relevant pulsational observables, namely periods, amplitudes and 
the complete topology of the
instability strip, where the blue boundaries for fundamental and first
overtone modes are connected to the position of the H and He
ionization regions within the pulsating envelope, whereas the red
boundaries are due to the quenching produced by convection, as well as the
variations of the relevant quantities (namely luminosity and
radial velocity)  along the pulsation cycle.  The capability
to reproduce light and radial velocity curves allows to directly
compare observed and predicted variations.  We also remind that, as
discussed in BCM00, in order to properly
reproduce all the morphological features exhibited by the first overtone
curves, one needs  models constructed by
assuming a vanishing overshooting efficiency in the regions where the
superadiabatic gradient attains negative values.  All the models
adopted in the present investigation are computed with this
assumption on the treatment of turbulent convection and include an
updated input physics 
(see BCM00 for details).  

\section{FIT OF THE OBSERVED LIGHT CURVES}

In order to reproduce the observed light curves of the selected 
RR Lyrae stars we adopted as input parameters the metal abundances and 
the  periods  
reported in Table \ref{t:table1}, and for each pulsator we computed isoperiodic model sequences 
(corresponding 
to the observed periodicity) for  the global metal abundance $Z$ inferred 
from the empirical $[Fe/H]$ 
value \footnote{We adopted the relation $\log{Z}=[Fe/H]-1.7$,  
neglecting the possible occurrence of $\alpha$ enhancement phenomena}. 
 Then for each model the theoretical $M_V$ {\it versus} Phase 
curve was overimposed 
to the data, and shifted vertically in magnitude to match the 
observed behaviour, 
by allowing at the same time 
the model input Z to vary within the
ranges permitted by the empirical determination. The necessary 
magnitude shift provides a direct evaluation
of the apparent distance modulus of the selected objects.

 Although the 
best fit solution is searched  over
 a wide range of stellar parameters,  
 the stellar masses were selected 
in order to cover at least the evolutionary  predictions 
for HB stars of the corresponding  metal 
abundances (see e.g. 
Cassisi et al. 1998, Pietrinferni et al. 2004, D'Antona
et al. 2002, Caloi \& D'Antona 2005,
Sweigart \& Catelan 1998, Catelan et al. 1998, VandenBerg et al. 2000). For each adopted stellar mass, 
the model luminosity and  
effective temperature were then varied in order to simultaneously reproduce 
the period and the light curve morphology in the $V$ band. To this 
purpose, for each  model, the predicted bolometric light curve was converted 
into the $B,V$ bands adopting the model atmospheres by Castelli, 
Gratton \& Kurucz (1997a,b).
When needed, the standard mixing length value ($l/H_p = 1.5$) 
was increased up to $l/H_p = 2.2$,
again in the range of the values adopted in current evolutionary computations.

The {\it best fit} models resulting from this investigation are shown 
in Figs. 1 and 2 for the fundamental mode and the first
overtone pulsators separately, and for both $V$ (left panels) and $B$ (right panels) 
light curves.
The corresponding intrinsic stellar parameters, namely the stellar mass, 
luminosity and effective temperature,  
are labelled for each pulsator (left panels) and summarized in Table \ref{t:table2}, along
with the input periods,  the adopted global metallicities $Z$, and mixing
length parameter choices. 
We find that models that best reproduce  the $V$ band 
light curves also show a good agreement with observations in the $B$ band.
 We also found that models corresponding to the spectroscopic metallicities 
(transformed to Z values) were generally  able to best fit the observed light
variations. Only in the case of stars \#5902 and \#2517 we had to respectively 
decrease and increase the input empirical metallicity 
to achieve a good reproduction of the observed light curves. However, these 
variations were always within the range allowed by the errors of the 
spectroscopic determinations.

The inferred apparent $V$ and $B$ distance moduli, $\mu_V$ and $\mu_B$, and their related 
uncertainties are
reported in 
Columns 9 and 10 of Table \ref{t:table2}, whereas the {\it pulsational} color 
excess $E(B-V)$ derived 
from the difference $\mu_B-\mu_V$ is given in Column 11. 
True individual $V$ distance 
moduli $\mu_0$ evaluated from the $\mu_V$ values in Column 9 of Table~\ref{t:table2} and  the 
adopted reddening values (see  Section 4.4)
assuming a standard extinction law $A_V=3.1*E(B-V)$, are given 
in Column 12. Finally, in Column 13 we provide the absolute visual
magnitudes derived from the model fitting, that is the intensity-averaged mean 
absolute magnitudes obtained from the best fit  model curves in the $V$ band. 
We note that, since these absolute magnitudes are intrinsic properties of 
the models,
 they only depend on  the accuracy of the model fitting and of the
 observed curves, 
 but are completely 
 independent of our knowledge of the  variable star reddening and distance.

As shown in Figs. 1 and 2, models are not able to completely reproduce
the rising branch of some of the fundamental mode pulsators. This is 
particularly true for the coolest pulsator in our sample, namely 
variable star V9660 for which the appearance in the model light curve of a 
{\it hump} before the maximum light, changes the slope of the rising branch and 
causes a delay of the phase of maximum itself. A similar effect is 
also present, although at lower extent, 
in the fitting of  V2249 and 
V2525. 
Analogous problems were found  by Castellani et al. (2002) to fit with 
current pulsation 
models the fundamental mode RR Lyrae stars at the FRE in the globular cluster
$\omega$ Cen.  

As for the first overtone pulsators,  the light curves are found to become 
close to pure sinusoids as the FOBE is approached (see the case of 
variable V2119).This behaviour holds for all the explored stellar masses 
and luminosity levels and might suggest the occurrence of some kind of 
degeneracy of the solution and the need for fixing a  M-L relation. However, 
since the range of effective temperature covered by almost pure sinusoidal 
light curves is quite narrow (100-150 K) at each luminosity level, 
by selecting the stellar mass in the range of the evolutionary predictions 
for the adopted metal abundance,  the best fit mass and  luminosity are 
constrained by the observed period and amplitude. 
We note, however, that the very short period,  the low apparent
luminosity, and the 
{\it sinusoidal} light curve of variable V2517 could also 
make it consistent with a fundamental mode $\delta$ Scuti model 
(with stellar mass in the range 1.5-2.0 $M_{\odot}$), according to the 
predictions by Bono et al. (1997). 

Finally, we point out that the mixing length
 parameter adopted to fit the observed light curves has the standard value 
$l/H_p=1.5$ for all the first overtone pulsators, whereas it ranges from 
1.85 to 2.2 for the fundamental mode variables, again in 
agreement with previous results from the model fitting technique.

\subsection{Uncertainties of the derived intrinsic quantities}  

In order to evaluate the uncertainties associated to the fitting
 procedure and their 
impact on the inferred intrinsic quantities, 
we have computed additional model sequences for a subsample 
of fundamental and first overtone pulsators, 
by varying the stellar parameters around the best fit values. 
 As a result of this procedure we find mean uncertainties of 
 $\delta \log{L/L_{\odot}} = 0.015 $, $\delta \log{T_e} = 25 K $ and
 $\delta \log{M/M_{\odot}} = 0.02 $. These uncertainties contribute an 
 error of $\sim$ 0.05 mag\footnote{This is an average value, the worst
 model fitting (star \#9660) having an uncertainty  0.06 mag.}  in the derived apparent 
 distance moduli $\mu_V$ and  $\mu_B$. Final errors on $\mu_V$ and  $\mu_B$ 
 should also include 
 the contributions due to the photometric uncertainties of the observed light 
curves.
 These are provided in Columns 5 and 8 of Table~\ref{t:table1}, and once 
 added in quadrature to the  0.05 mag uncertainty of the fitting 
procedure lead to the
 total errors associated to the individual $\mu_V$, $\mu_B$ values in 
Columns 9 and 10 of Table~\ref{t:table2}.

 We are aware that the present results hold in the context of  the adopted 
 theoretical 
scenario,  namely Bono et al. (2003), and Marconi et al. (2003) pulsation 
models. 
However, we note that Bono, Marconi \& Stellingwerf (1999) 
performed a direct comparison with WAS pulsation models.   
By adopting the same input parameters of WAS
for constructing the light curve of the bump Cepheid HV905, they found 
that predictions by the two hydrocodes, although based on 
different physical assumptions concerning the input
physics and the pulsation/convection interaction, are in reasonable agreement. 

Errors on the {\it pulsational} color excess E(B-V) in  
Table~\ref{t:table2} are computed as the propagation of 
uncertainties in the  $\mu_V$ and  $\mu_B$ values (Columns 9 and 10 of 
Table~\ref{t:table2}). Admittedly, they are rather large. Still 
the {\it pulsational} reddenings derived for these LMC variables are
very reasonable values, and agree well with estimates from
other more robust methods (e.g. C03, see Section 4.4), thus providing anyway
a sanity check both of other reddening determinations and 
of the reliability of our model fitting procedure. 

Finally, errors on the true $V$ distance moduli
$\mu_0$ are the sum in quadrature of 
3.1 times the uncertainty in the adopted
reddenings (see Section 4.4) and of the errors in the apparent
$\mu_V$'s in Column 9 of Table~\ref{t:table2}. Uncertainties in the $M_V$'s 
instead only depend on  
the uncertainty of the best fitting procedure (0.05 mag), and 
on the photometric uncertainties of the observed $V$ light curves 
(Column 5 of Table~\ref{t:table1}). 

We explicitly note that the smaller errors 
of the  $M_V$ values derived from the model fitting compared to 
the errors of the $\mu_0$ values reflect the strength of  
a technique which allows to directly estimate the 
intrinsic quantity, namely the absolute magnitude of the variable star,
without  relying upon the knowledge of the reddening,
by far one of the most uncertain and controversial quantities 
in the LMC.

\subsection{Comparison with the intrinsic parameters predicted from the evolutionary
models}
 
Stellar masses and luminosities derived from the model fitting 
are generally consistent, within the uncertainties on the model fitting quoted in Sect. 4.1, with 
the values predicted by stellar evolution models of the proper metal 
abundances. This is shown in Table~\ref{t:table3} where 
the comparison is made with the evolutionary Zero Age Horizontal Branch (ZAHB)
 models 
 of  Cassisi et al. (1998) and Pietrinferni et al. (2004), 
 D'Antona et al. (2002) and Caloi \& D'Antona (2005),  
 Sweigart \& Catelan (1998) and Catelan et al. (1998), and finally VandenBerg et al. (2000).
  The comparison among estimates given by different authors helps us to 
 quantify the uncertainty of current evolutionary predictions.
 For some of the variables we find that the pulsation model fitting provides significantly smaller 
 masses and/or higher luminosity levels than  all the considered ZAHB  
 models, thus suggesting possible  evolutionary effects.  
 This is indeed the case for the metal-poor variable stars V5902, V7620, V7477, V2024, and for 
 the metal-intermediate stars V2249, V4179 and V27697. For the relatively metal-rich variable 2517 
 our pulsational luminosity is slightly fainter than the evolutionary 
 predictions, but consistent with Catelan et al. estimate, within the error. 
 Moreover, as already mentioned, the model metallicity for this star was 
 slightly increased with respect to the spectroscopic value, within the 
 empirical uncertainty. Indeed, a best fit solution was also found for the mean 
 spectroscopic value (Z=0.002), however in this case the intrinsic model parameters 
 were not in agreement with the evolutionary predictions. 
 Finally, as noticed in Sect. 4, this star, given the very short period and low apparent luminosity, 
 could also be consistent with 
 a relatively massive fundamental mode $\delta$ Scuti pulsator.


\subsection{Comparison with the intrinsic parameters from the Fourier
decomposition of the light curves}

 In Table \ref{t:table4} we have summarized masses, luminosities, effective 
temperatures, and absolute magnitudes derived from the present study and, for
comparison,  the same quantities as estimated by DF04 from the Fourier 
parameters
of the visual light curves for the variables in the present sample
for which this analysis was possible, namely 5 RRab's that 
satisfy Jurcsik \& Kov\`acs (1996) {\it compatibility conditions} 
 (see details in Section 6 of DF04).

We note that while masses estimated from the two methods agree within a few
hundreds, luminosities and effective temperatures from the model fitting are
on average systematically larger, respectively by 0.07 dex and 200 K, than the 
corresponding quantities 
from the Fourier analysis. These systematic effects seem to suggest the 
occurrence of problems with the calibration of Fourier parameters, 
making them still not reliable (see also the discussion in Cacciari et al. 2004).

Moreover, the absolute magnitudes from the model fitting
span an interval about 0.13 mag larger than those from the Fourier parameters,
a similar result was also found by DF04.

\subsection{Comparison of reddening estimates}

The amount of reddening affecting the LMC still 
remains poorly known since estimates by different techniques and
authors have often provided controversial results that do not agree
to each other (see  U99, Zaritsky 1999, C03, A04, Zaritsky et al.  
 2004). The LMC color excess is also found to
vary by large amounts from one region of the galaxy to the other, thus 
requiring local estimates that are to be preferred to the use of 
average values over large areas.
The {\it pulsational} reddenings, being derived for each star individually,
indeed provide estimates of the {\it local} reddening in the region containing
the variable, that can be directly compared to other 
{\it local} estimates of the $E(B-V)$.

In Table~\ref{t:table5} the {\it pulsational} reddenings derived  
from the model fittings are compared with estimates 
obtained for the same stars or LMC regions 
by C03, U99, and A04 using different techniques.
In particular, reddenings labelled ``strip'' 
(see Column 4 of Table~\ref{t:table5}) 
were obtained by C03 from the comparison of the colors of 
the edges of the RR Lyrae instability strip defined by 
the variables contained in the two 
LMC fields they observed,
with those of the low reddening globular cluster M3.
They determined two distinct values, that apply to the
variables contained in each given field, separately.
Reddenings labelled ``Sturch'' (see Column 5 of Table~\ref{t:table5})
are ``individual'' 
estimates derived by 
C03 using Sturch (1966) method, which is based on the color at
minimum light of the fundamental mode RR Lyrae stars.
U99 is the average reddening measured from variations in the 
$I$ luminosity of clump stars located in 
OGLE-II field LMC\_SC21. This field partially overlaps to C03 Field A, and 
according to Table~14 of DF04, two of the variable stars
analyzed here (namely the RRc stars \#2119 and \#27697) fall into
the common region. 

 Finally, A04 is the mean color excess estimated
from 330 first overtone RR Lyrae stars spread over 16 MACHO fields 
close to the LMC bar. 
We only have the RRc star \#8837 in common with 
A04 sample (see Table~12 in that paper).

As discussed in Section~4, notwithstanding their large uncertainties,
the {\it pulsational} reddenings provide a consistency
check of the color excesses derived by other techniques.
In this respect we note that they agree well with the
empirical values by C03 while, on average, are about 
0.02-0.03 mag  smaller than U99 and A04 estimates.  
However, this systematic difference is not surprising given the rather
clumpy reddening of the LMC, the larger areas covered by the 
reddening indicators used in both U99\footnote{The common area 
with C03 Field A is less than 10\% of OGLE-II field LMC\_SC21,
that extents both more inside and outside the LMC bar.} and A04, 
and their locations on or  closer to the LMC bar than in C03.

In the following we will adopt for each of our targets the weighted average of
the {\it pulsational}, C03-strip, and C03-Sturch reddenings 
(values in Columns 3,4 and 5 of Table~\ref{t:table5}).
These average values are provided in Column 8 with their related uncertainties
that are simply the standard deviations of the 
weighted means.  They will be used together with the apparent distance moduli 
$\mu_V$ in Table~\ref{t:table2} to derive true distance moduli $\mu_0$ for each
of our targets (see Section 5).

\section{THE DISTANCE TO THE LMC}



True distance moduli $\mu_0$ for our program stars were computed from 
the apparent moduli
in Column 9 of Table~\ref{t:table2} and the reddenings in Column 
8 of Table~\ref{t:table5} using the standard extinction law 
($A_V$=3.1$\times E(B-V)$). They  
are provided in Column 12 of Table~\ref{t:table2}.
Errors in these $\mu_0$ values are the sum in quadrature of 
the uncertainties in the apparent distance moduli
$\mu_V$  and 3.1 times the uncertainty of the reddenings.     
By performing a weighted mean of these $\mu_0$ values 
 we obtain a final estimate 
for the true distance modulus of the LMC of
$<\mu_{0}>$= 18.54 $\pm$ 0.02 ($\sigma$=0.09, standard deviation about
the average of
 the 14 RR Lyrae stars).
 The 0.09 mag dispersion of this average value is fully accounted for by 
uncertainties in the reddening and
model fitting technique, implying that the distance moduli derived for individual
objects are consistent within the respective error bars.   

The derived LMC average distance modulus is  well in 
the range of the most recent evaluations in the literature 
that all prefer the {\it long} distance scale (see Walker 2003, C03, Alves 
2004, and reference therein),  agrees well  
with the results of the light curve fitting of two Classical 
Cepheids  in the LMC by BCM02,  and is in excellent agreement with 
Keller \& Wood (2002) LMC distance modulus  from the model fitting of
20 LMC {\it bump} Cepheids. This excellent agreement between  
Population I and II
distance indicators in the LMC is very rewarding in light of the longstanding 
controversy existing in the past  
on the distance to the LMC 
based on these two independent indicators.  Moreover the agreement with
Keller \& Wood determination, which is based on a different pulsation code 
and relies
on different physical and numerical assumptions, supports the soundness 
and reliability of our results.

We also point out that a different choice for the reddening, namely adoption
of U99 and A04 average values ($E(B-V)$=0.142 mag) would give 
$<\mu_{0}>$=18.46$\pm 0.02$, 
again in the range of the ``long'' scale and 
at odds with methods  favoring much shorter distance 
moduli (in the range from $\sim$18.2 to $\sim$18.3 mag) for the
LMC (e.g. Fernley et al. 1998a, Udalski 2000, Popowski 2000, Dambis 2004, 
Rastorguev, Dambis, \& Zabolotskikh 2004).

%
%
%

%
%
%

\subsection{The RR Lyrae luminosity-metallicity relation}

 HB and RR Lyrae stars are known to follow a luminosity-metallicity
relation generally considered to be of linear form. The slope of this 
relation is still matter of debate with values in the range from 0.30 
(Sandage 1993) to 0.18-0.20 mag/dex (Caloi et al. 1997, Cassisi et al. 1998, 
Fernley et al. 1998b, C03, G04, Rich et al. 2001, 2005). 
There are also empirical and theoretical
evidences for a non-linearity of the relation 
followed by the Galactic globular clusters (Caputo et al. 2000, 
Rey et al. 2000), however there is no clear proof 
of such a non-linearity in the behaviour of the Galactic (Fernley
 et al. 1998b) and LMC (C03, G04) field RR Lyrae stars, as well as 
 in the M31 globular clusters (Rich et al.
2001, 2005).  At fixed metal abundance there is also an intrinsic spread in the HB 
luminosity, due to evolutionary effects, 
(see e.g. Sandage 1990). 
Such evolutionary effects are also predicted by  evolutionary and synthetic horizontal branch 
computations (see e.g. Lee, Demarque \& Zinn 1990, Caputo et al. 1993, 
Caloi et al. 1997, Cassisi et al. 2004 and references therein).
A fairly large number of objects should be considered in order to reduce the 
impact of the evolutionary effects, since evolution off the ZAHB
  can affect the
slope of the luminosity-metallicity relation.

C03 and G04 have recently derived the slope of the luminosity-metallicity of the
LMC RR Lyrae stars using a fairly significant 
large sample of variables (98 stars) based on their homogeneous and 
accurate photometric and spectroscopic datasets for these stars.
Since RR Lyrae stars in the LMC can in  the first approximation 
all be considered at the same distance from us, C03 and G04 used directly
the dereddened apparent visual magnitudes of the variables 
without any assumption on their absolute magnitudes, that might be affected
by problems of zero point in the distance scale.
Further advantages of C03 and G04 approach were the significant statistics of
their sample, and the large number of objects at intermediate metal abundance:
the metal distribution of C03 and G04 sample peaks at [Fe/H]=$-$1.48 dex, with
the bulk of stars (66 objects) in the metallicity bin $-1.7<$[Fe/H]$<-$1.3.  Both
these characteristics allow minimization of  the effects on the slope of the 
intrinsic spread in luminosity
of the RR Lyrae stars due to their evolution off the ZAHB.   
They derived a rather mild slope of 0.214 $\pm$ 0.047 mag/dex in agreement with 
results from the Baade-Wesselink method of 28 Galactic field RR Lyrae stars
($\Delta M_V/[Fe/H]$=0.20 $\pm$ 0.04 mag/dex, Fernley et al. 1998b), and from the
HB luminosity of 20 globular clusters in the Andromeda galaxy 
($\Delta M_V/[Fe/H]$=0.20 mag/dex $\pm$0.09, Rich et al. 2005).

We have combined the $<M_V>$ values derived from the model fitting 
(see Column 13 of Table~\ref{t:table2}) 
with the metallicities used in the
fitting (see Column 4 of Table~\ref{t:table2}) to determine the slope
of the luminosity-metallicity relation defined by the 
14 LMC RR Lyrae stars in our sample.
A simple linear fit provides a slope of  $\Delta M_V/[Fe/H]$=0.34 mag/dex 
using all objects and of 0.28 mag/dex if star \# 2517, the object with rather
unusually faint absolute magnitude, is discarded. The same linear regression 
using the corresponding dereddened apparent magnitudes $V_0$ (obtained from 
the average $<V>$ values in Column 4 of Table~\ref{t:table1} corrected for
the reddenings in Column 8 of Table~\ref{t:table5} assuming the standard extinction 
law) would give  $\Delta M_V/[Fe/H]$=0.31 and 0.24, respectively.
All these values are higher than the slope derived by C03 and G04. However, 
since  the  $V_0$ {\it versus} [Fe/H] fit also provides steeper slopes 
compared to C03 and G04 value, this demonstrates that the difference is due to  
the sample selection,  and that it does not arise from errors in the 
absolute magnitudes obtained by the model 
fitting.  Indeed, it should be kept in mind that the 14 LMC RR Lyrae stars analyzed 
in this paper represent less than 15\%  of C03 and G04 sample, and that although the 
peak of the fit metallicities of our RR Lyrae subsample 
([Fe/H]$_{fit}$=$-$1.53) is very close to the average metallicity of C03 and 
G04 total sample,  our metallicity distribution is more unbalanced 
towards low metal abundances. Moreover,the comparison with the ZAHB evolutionary 
predictions indicates the presence of evolutionary effects for almost all 
the metal poor objects in our sample,  and these effects are expected to cause a steepening of
the slope.
Thus the higher slopes found here are simply the result of both
the poorer statistics and the higher incidence of  evolutionary effects
in our sample compared to C03 and G04 one.  

\section{SUMMARY AND FINAL REMARKS}

We have fitted the nonlinear convective pulsation models by Bono et al. (2003) and 
Marconi et al. (2003) to 14 LMC RR Lyrae stars 
with accurate photometry, metal abundances, and reddening estimates 
by DF04, G024 and C03.
This is the first time that the model fitting technique is applied to a
significantly large number of RR Lyrae stars within the same stellar system, thus
providing an important assessment of the predictive capabilities of the adopted
theoretical pulsation models and, at the same time, a new independent estimate of the 
Population II
distance to this fundamental fist step of the astronomical distance ladder.

We have obtained $\mu_0(LMC)$=18.54 $\pm$0.02 in very good agreement with the
LMC ``long'' astronomical distance (Walker 2003, C03, Alves 2004).

Masses, luminosities, and effective temperatures derived from to the model
fitting are in satisfactory agreement with the predictions of theoretical HB models, 
as derived from various authors in the literature. We notice that 
such a detailed and punctual comparison between  the results of pulsation model fitting
and the evolutionary expectations had never been performed in the previous applications 
of the method to RR Lyrae stars.
 
The {\it pulsational} reddenings are in good agreement with values in C03, 
 thus further supporting the estimates in that paper, and the soundness of the
present approach.

Finally we remind that our final result for the LMC distance modulus is in excellent 
agreement with the estimate provided by Keller \& Wood (2002) on the basis of a similar 
method applied to 20 Classical Cepheids in the same stellar system. This occurrence 
suggests that the results of the model fitting technique are only marginally dependent on
the adopted pulsation code, as well as on the physical and numerical assumptions 
in the model computations.  It also shows that there is no discrepancy between the 
Population I and Population II 
distance scales to the Large Magellanic Cloud.

\acknowledgments
We warmly thank Santi Cassisi, Marcio Catelan and Don Vandenberg for providing us 
their ZAHB models in electronic forms.  A special thanks goes to Vittoria Caloi and Franca D'Antona
for sending us their evolutionary models in advance of publication and for 
specifically computing some of the ZAHBs used in the comparison 
with the evolutionary models. It is a pleasure to thank Franco Zavatti 
for many enlightening discussions on the regression line routines used 
in astronomy,  and the anonymous referee for useful comments.
Financial support for this study was provided by 
INAF Progetti di Ricerca di Interesse Nazionale under the scientific project
``Stars and Clusters as Tracers of the LMC Structure and Evolution'' (P.I.: 
G. Bertelli), and
by MIUR, under the scientific projects ``Stellar Populations in the Local Group'' (P.I.: Monica Tosi) and  ``Continuity and Discontinuity in the 
Milky Way Formation'' (P.I.: Raffaele Gratton).


\clearpage

\begin{figure}
\includegraphics[width=15cm]{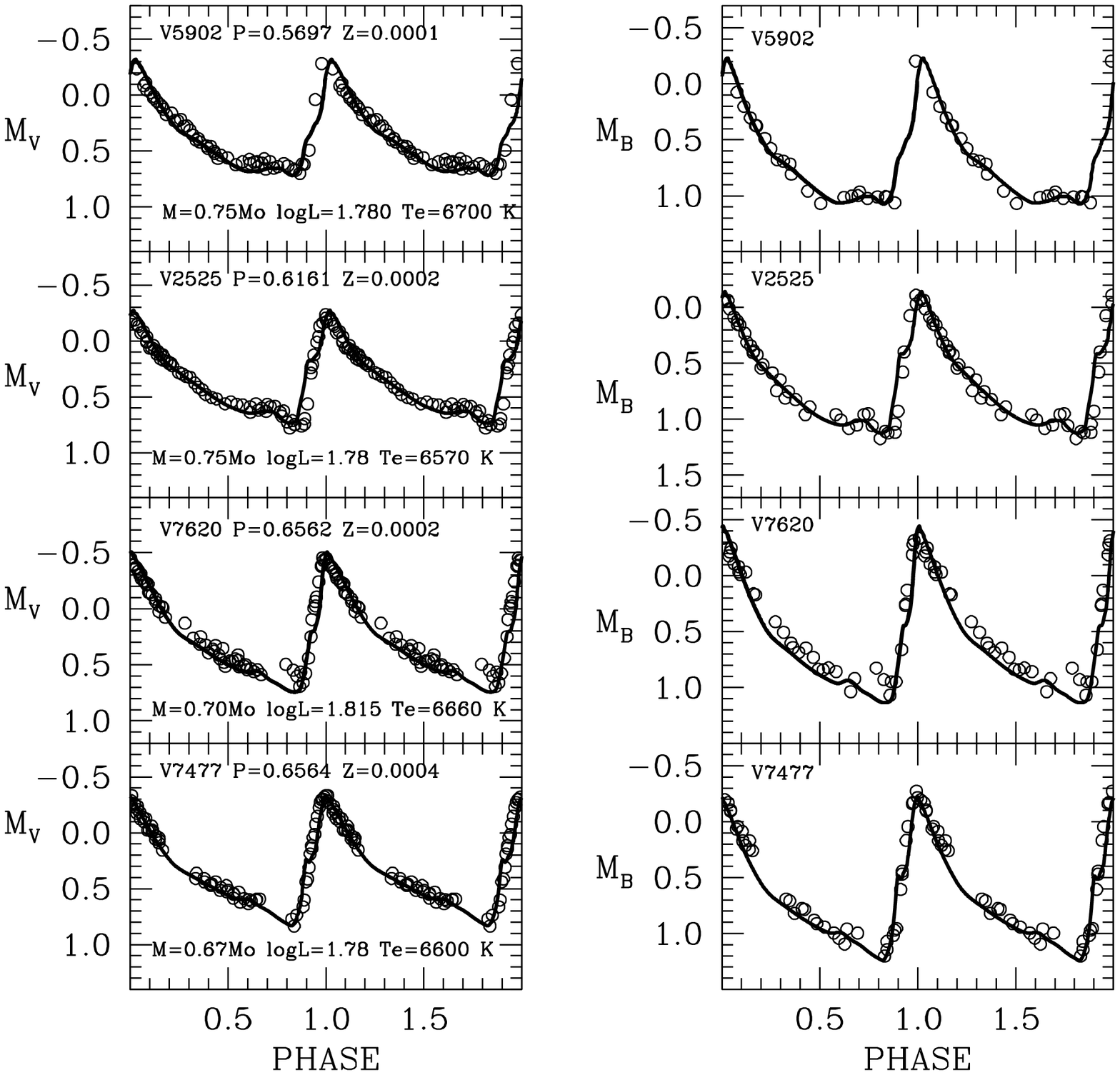}
\figcaption{
Results from the theoretical modelling of the $V$ (left panels)
 and $B$ (right panels) light curves of the fundamental mode pulsators in our
 sample. Stars are ordered by increasing metallicity. For each variable we indicate
 the star identifier and period according to DF04, the global metal 
 abundance $Z$ inferred from the [Fe/H] values in 
 G04 or DF04, and the mass, luminosity 
 and effective temperature obtained from the fit.}
\label{f:figure1}
\end{figure}

\clearpage
\begin{figure}
\includegraphics[width=15cm]{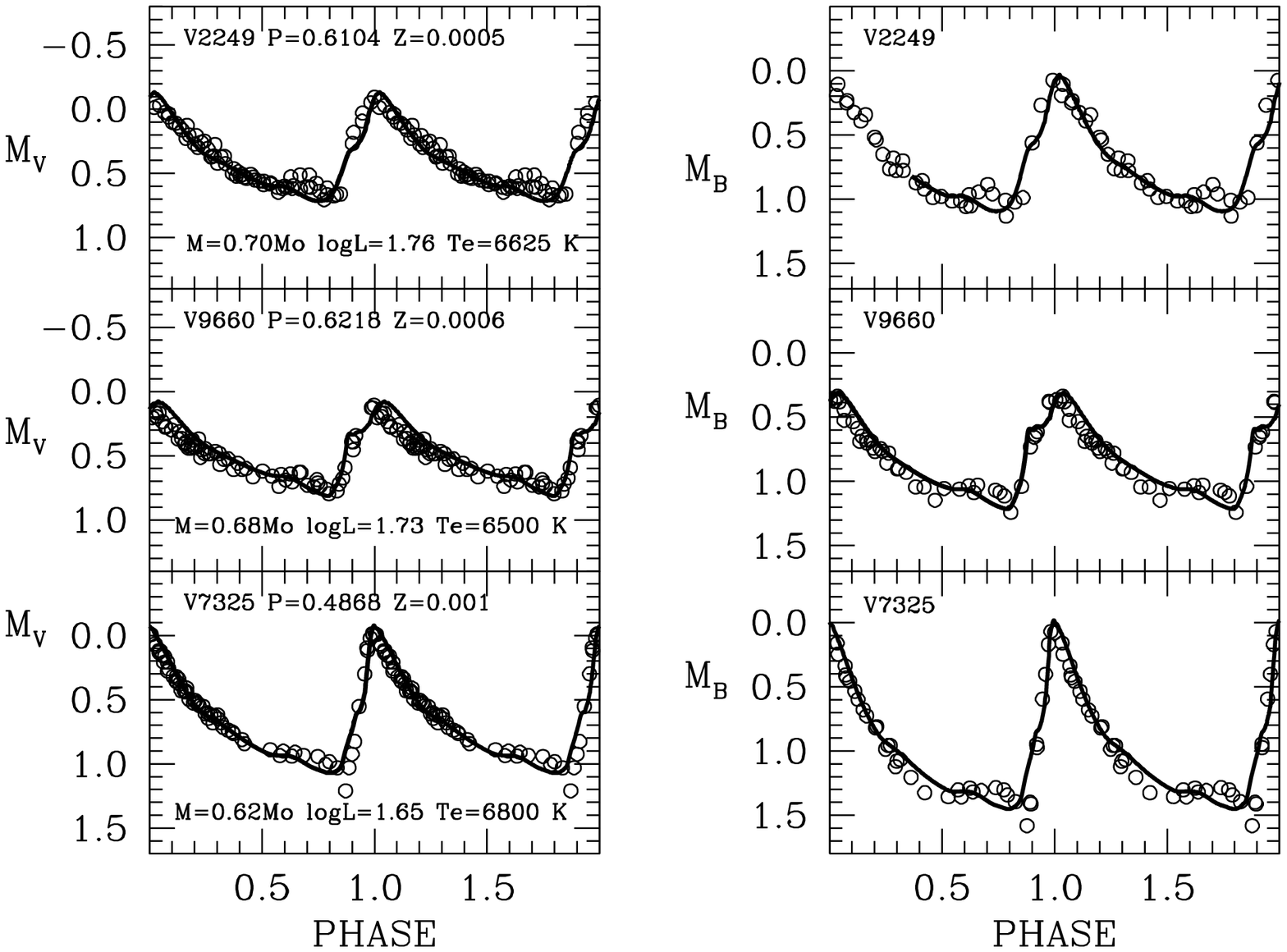}

{Fig. 1 -- continued --}
\label{f:figure1b}
\end{figure}

\clearpage
\begin{figure}
\includegraphics[width=15cm]{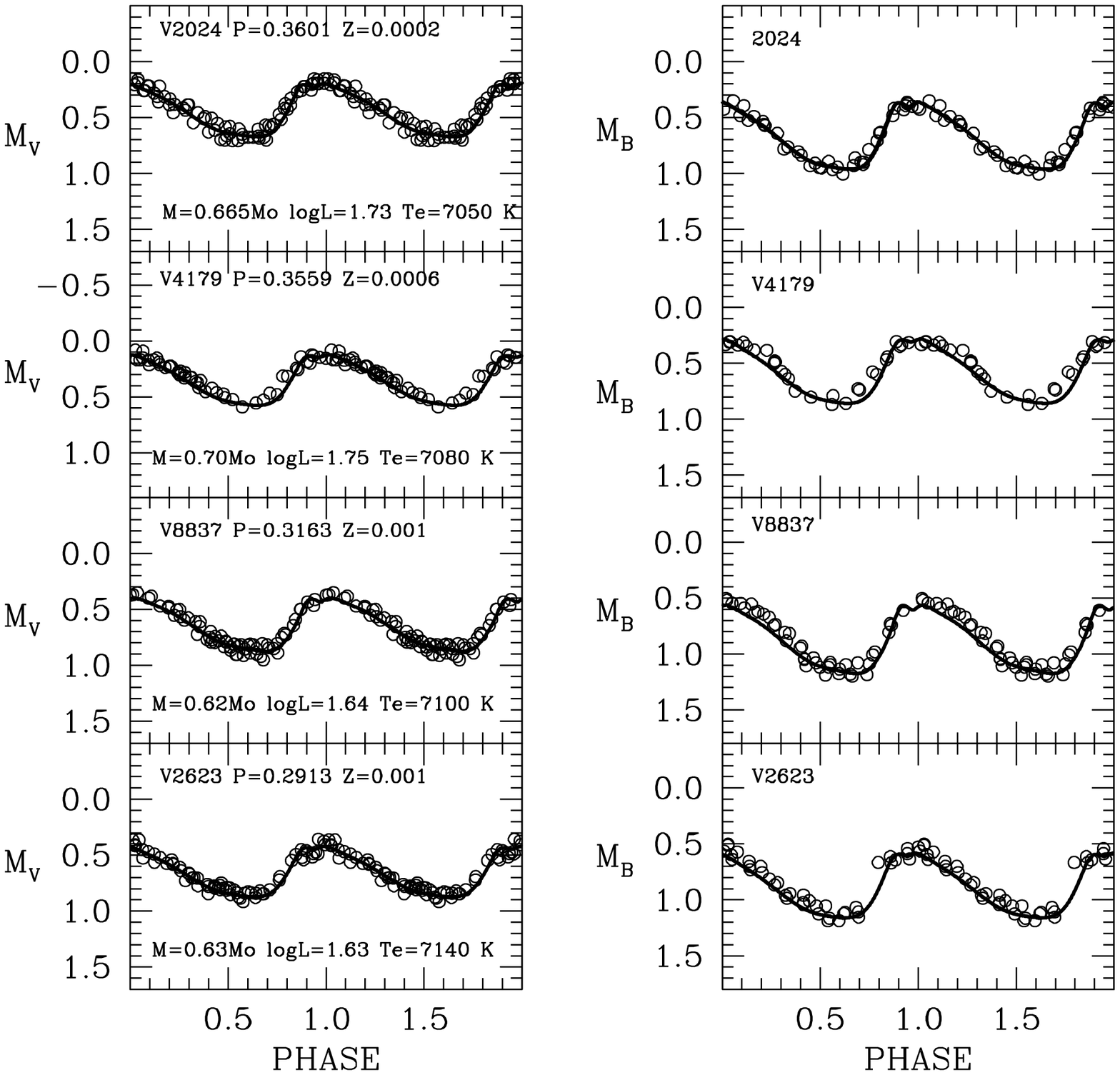}
\figcaption{Results from the theoretical modelling of the $V$ (left panels)
 and $B$ (right panels) light curves of the first overtone pulsators in our
 sample. Stars are ordered by increasing metallicity. 
 For each variable we indicate
 the star identifier and period according to DF04, the global metal abundance 
 $Z$ inferred from the [Fe/H] values in 
 G04 or DF04, and the mass, luminosity and effective temperature obtained from the
 fit.}
\label{f:figure2}
\end{figure}

\clearpage
\begin{figure}
\includegraphics[width=15cm]{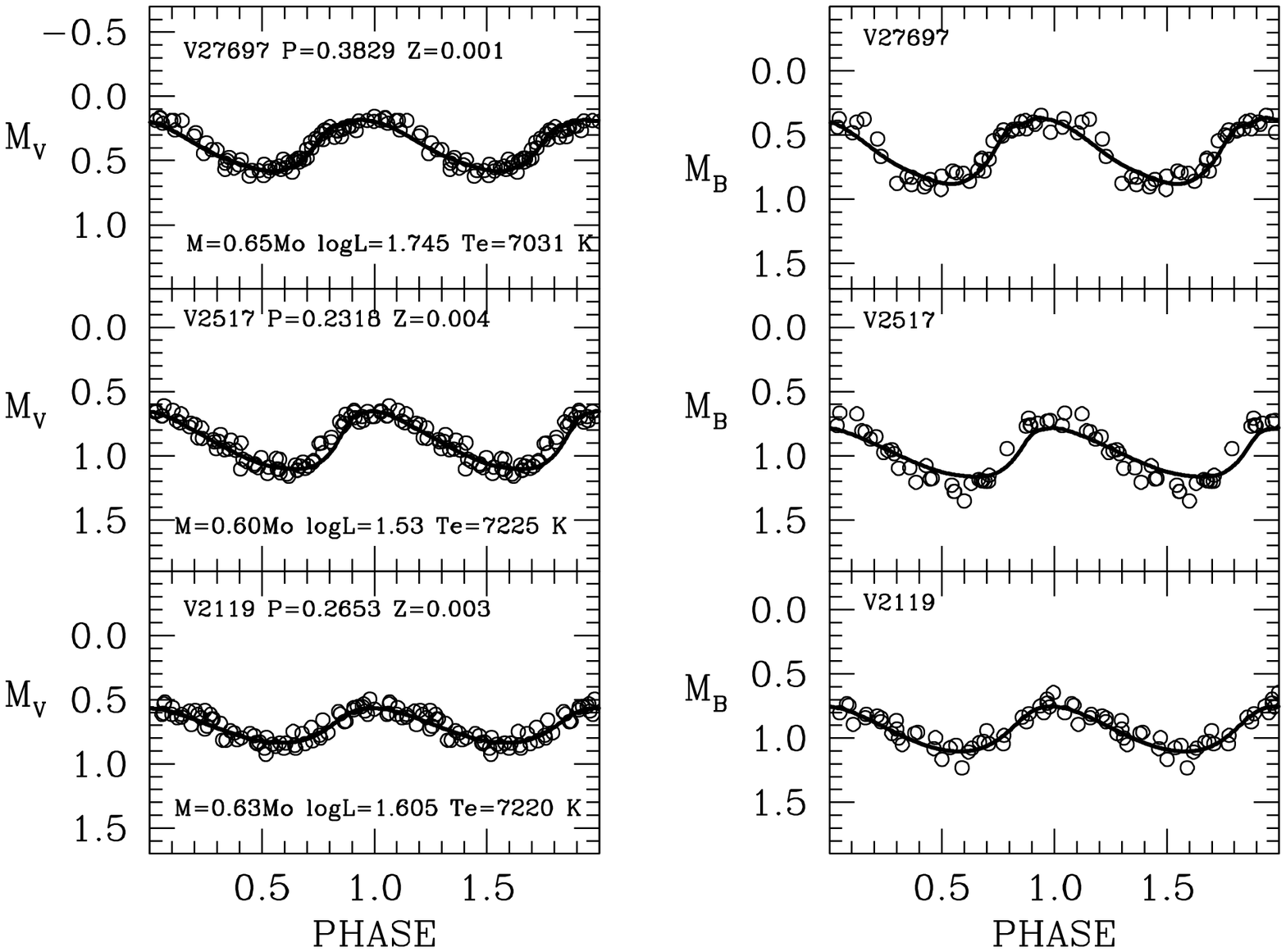}

{Fig. 2 -- continued --}
\label{f:figure2b}
\end{figure}


\clearpage
\begin{table*}[ht]
\caption{Observed characteristics of the program stars}
\vspace{0.5cm}

\scriptsize
\begin{tabular}{rcccccccccccccl}
\tableline \tableline
Star$^a$ &type$^a$ & P$^a$   & $<V>^a$ & $\sigma_V$ &N$_p$ & $<B>^a$& $\sigma_B$ &N$_p$&A$_V^a$& A$_B^a$ &[Fe/H]$^b$\\ 
         &         & day     & mag    &  mag    &  $V$    & mag     & mag     & $B$  &mag  &  mag    &          \\
\tableline
5902 & ab & 0.570 & 19.121 & 0.033 &54 & 19.472 & 0.037 &23 & 1.015& 1.282 & $-2.12 \pm 0.11 $\\ 
2525 & ab & 0.616 & 19.340 & 0.036 &69 & 19.764 & 0.063 &41 & 0.991& 1.272 & $-2.06 \pm 0.14 $\\ 
7620 & ab & 0.656 & 19.079 & 0.040 &70 & 19.409 & 0.067 &37 & 1.071& 1.366 & $-2.05 \pm 0.12 $\\ 
7477 & ab & 0.656 & 19.183 & 0.033 &69 & 19.552 & 0.057 &40 & 1.108& 1.371 & $-1.67 \pm 0.28 $\\ 
2249 & ab & 0.610 & 19.346 & 0.041 &69 & 19.775 & 0.052 &35 & 0.747& 0.987 & $-1.56 \pm 0.15 $\\ 
9660 & ab & 0.622 & 19.392 & 0.038 &67 & 19.862 & 0.054 &41 & 0.669& 0.811 & $-1.50 \pm 0.20 $\\ 
7325 & ab & 0.487 & 19.435 & 0.035 &68 & 19.845 & 0.054 &40 & 1.131& 1.449 & $-1.28 \pm 0.09 $\\ 
2024 & ~c & 0.360 & 19.500 & 0.048 &72 & 19.876 & 0.059 &40 & 0.509& 0.606 & $-1.92 \pm 0.14 $\\ 
4179 &~c  & 0.356 & 19.173 & 0.037 &50 & 19.502 & 0.048 &29 & 0.438& 0.560 & $-1.53 \pm 0.27 $\\ 
8837 &~c  & 0.316 & 19.566 & 0.045 &64 & 19.905 & 0.059 &41 & 0.501& 0.631 & $-1.52 \pm 0.22 $\\ 
2623 &~c  & 0.291 & 19.368 & 0.043 &65 & 19.631 & 0.064 &40 & 0.441& 0.595 & $-1.34 \pm 0.30 $\\ 
27697&~c  & 0.383 & 19.166 & 0.041 &66 & 19.541 & 0.064 &39 & 0.396& 0.471 & $-1.33 \pm 0.25 $\\ 
2517 & ~c & 0.232 & 19.695 & 0.049 &69 & 19.925 & 0.052 &36 & 0.452& 0.555 & $-0.92 \pm 0.33 $\\
2119 & ~c & 0.265 & 19.659 & 0.052 &64 & 19.986 & 0.071 &39 & 0.297& 0.354 & $-0.79 \pm 0.26 $\\ 

\tableline
\end{tabular}
\label{t:table1}
\normalsize
\vspace{0.5cm}

$^a$ Star identifier, type, period, intensity-averaged mean magnitudes with the related uncertainties, number 
of data points, and amplitudes of the light 
variations are
from DF04

$^b$ Metallicities [Fe/H] are on G04 metallicity scale.

\end{table*}

\begin{table*}[ht]
\caption{Intrinsic parameters, distance moduli, reddenings and absolute
magnitudes derived from
the modelling of the light curves
}
\vspace{0.5cm}

\tiny
\begin{tabular}{rccllcccccccc}
\tableline \tableline
Star&type& P   &~~~$Z^a$&$l/H_p$& $M/M_{\odot}$ & $\log{L/L_{\odot}}$ & $T_e$ & $\mu_V$ & $\mu_B$ & $E(B-V)$ &
$\mu_0$& $<M_V>$ \\ 
    &    &  day&        &       &               &                     & K     & mag     &  mag    &  mag     &  
mag    &  mag    \\ 
\tableline
5902 & ab & 0.570&0.0001 &2.1     &0.75 &1.78 &6700 & 18.81$\pm$0.060 & 18.87$\pm$0.062 & 0.06$\pm$0.086& 18.519$\pm$0.080& 0.347$\pm 0.060$\\
2525 & ab & 0.616&0.0002 &1.9     &0.75 &1.78 &6750 & 18.99$\pm$0.062 & 19.12$\pm$0.080 & 0.13$\pm$0.101& 18.562$\pm$0.130& 0.336$\pm 0.062$\\  
7620 & ab & 0.656&0.0002 &1.89    &0.70 &1.815&6660 & 18.87$\pm$0.064 & 18.97$\pm$0.084 & 0.10$\pm$0.106& 18.579$\pm$0.074& 0.252$\pm 0.064$\\  
7477 & ab & 0.656&0.0004 &1.85    &0.67 &1.78 &6600 & 18.89$\pm$0.060 & 18.99$\pm$0.076 & 0.10$\pm$0.096& 18.515$\pm$0.069& 0.336$\pm 0.060$\\  
2249 & ab & 0.610&0.0005 &2.2     &0.70 &1.76 &6625 & 18.99$\pm$0.065 & 19.12$\pm$0.072 & 0.13$\pm$0.097& 18.658$\pm$0.121& 0.392$\pm 0.065$\\  
9660 & ab & 0.622&0.0006 &2.0     &0.68 &1.73 &6500 & 18.89$\pm$0.063 & 19.03$\pm$0.074 & 0.14$\pm$0.097& 18.549$\pm$0.073& 0.467$\pm 0.063$\\  
7325 & ab & 0.487&0.001  &2.0     &0.62 &1.65 &6800 & 18.79$\pm$0.061 & 18.89$\pm$0.074 & 0.10$\pm$0.096& 18.437$\pm$0.062& 0.640$\pm 0.061$\\  
2024 & ~c & 0.360&0.0002 & 1.5    &0.665&1.73 &7050 & 19.08$\pm$0.069 & 19.21$\pm$0.077 & 0.13$\pm$0.103& 18.720$\pm$0.076& 0.435$\pm 0.069$\\  
4179 & ~c & 0.356&0.0006 & 1.5    &0.70 &1.75 &7080 & 18.85$\pm$0.062 & 18.96$\pm$0.069 & 0.11$\pm$0.093& 18.580$\pm$0.082& 0.348$\pm 0.062$\\  
8837 & ~c & 0.316&0.001  & 1.5    &0.62 &1.64 &7100 & 18.95$\pm$0.067 & 19.06$\pm$0.077 & 0.11$\pm$0.102& 18.590$\pm$0.068& 0.637$\pm 0.067$\\  
2623 & ~c & 0.291&0.001  & 1.5    &0.63 &1.63 &7140 & 18.73$\pm$0.066 & 18.80$\pm$0.081 & 0.07$\pm$0.104& 18.374$\pm$0.122& 0.652$\pm 0.066$\\ 
27697& ~c & 0.383&0.001  & 1.5    &0.65 &1.745&7031 & 18.80$\pm$0.065 & 18.93$\pm$0.081 & 0.13$\pm$0.104& 18.440$\pm$0.072& 0.376$\pm 0.065$\\
2517 & ~c & 0.232&0.004  & 1.5    &0.60 &1.53 &7225 & 18.825$\pm$0.070& 18.965$\pm$0.072& 0.14$\pm$0.100& 18.552$\pm$0.137& 0.889$\pm 0.070$\\
2119 & ~c & 0.265&0.003  & 1.5    &0.63 &1.605&7220 & 18.96$\pm$0.072 & 19.06$\pm$0.087 & 0.10$\pm$0.113& 18.600$\pm$0.080& 0.705$\pm 0.072$\\ 
\tableline
\end{tabular}
\label{t:table2}
\normalsize
\vspace{0.5cm}

$^a$ $Z$ values were derived from the observed metal abundances [Fe/H]
in Table~1 using the relation $\log{Z}=[Fe/H]-1.7$



\end{table*}

\begin{table*}[ht]
\caption{Comparison with the evolutionary masses and luminosities predicted 
by different sets of ZAHB models}
\vspace{0.5cm}

\tiny
\begin{tabular}{rcccccccccccccc}
\tableline \tableline
Star&type& P   & $Z$    & $\log{T_e}$ & $M$ & $M$ & $M$ &$M$ &$M$ &$\log{L}$ & $\log{L}$ & $\log{L}$ & $\log{L}$ & $\log{L}$ \\ 
    &    & day &        & Puls.  &  Puls.   &  C98/P04          & DC02/05     & SC98     & V00    &  Puls.      
&C98& DC02/05& SC98 & V00\\
\tableline          
5902 & ab& 0.570& 0.0001 &3.826& 0.75 & 0.84 & 0.84 & $-$ & 0.80 & 1.78 &1.79 & 1.79 & $-$ & 1.75 \\
2525 & ab& 0.616& 0.0002 &3.818& 0.75 & 0.76 & 0.77 & $-$ & 0.74 &1.78 &1.75 & 1.75 & $-$ & 1.72 \\ 
7620 & ab& 0.656& 0.0002 &3.823& 0.70 & 0.76 & 0.76 & $-$ & 0.74 & 1.815 &1.75 &1.75 & $-$ & 1.72\\ 
7477 & ab& 0.656& 0.0004 &3.819& 0.67 & $-$ & $-$ & $-$ & 0.69 & 1.78 & & & $-$ & 1.70\\ 
2249 & ab& 0.610& 0.0005 &3.821& 0.70 &  $-$    &$-$ & 0.68 & 0.67  &1.76 &$-$ &$-$ &1.67& 1.68  \\
9660 & ab& 0.622& 0.0006 &3.813& 0.68 &  $-$  & $-$ &$-$ & 0.66 &1.73 &$-$ &$-$ & $-$ & 1.68 \\ 
7325 & ab& 0.487& 0.001  &3.833& 0.62 & 0.66 & 0.65 & 0.64  & 0.63 &1.65 & 1.69 & 1.68 & 1.63 & 1.65\\ 
2024 & c & 0.360& 0.0002 &3.848& 0.665& 0.75 & 0.75 &$-$ & 0.72 &1.73 &1.73 & 1.74 & $-$ & 1.71\\ 
4179 & c & 0.356& 0.0006 &3.850& 0.70 & $-$     &$-$ & $-$ &0.65 &1.75 &$-$ &$-$ & $-$ & 1.67\\ 
8837 & c & 0.316& 0.001  &3.851& 0.62 & 0.65 &0.65 & 0.65  & 0.63 &1.64 &1.68 &1.67 & 1.63 & 1.65 \\ 
2623 & c & 0.291& 0.001  &3.854& 0.63 & 0.65 & 0.645 & 0.65  & 0.63 &1.63 &1.68 & 1.67 & 1.63 & 1.65\\ 
27697& c & 0.383& 0.001  &3.847& 0.65 & 0.66 & 0.645 & 0.65 & 0.63  &1.74 &1.68 & 1.67 & 1.63 & 1.65\\ 
2517 & c & 0.232& 0.004  &3.859& 0.60 & 0.59    & 0.59 &  0.59 & 0.575 &1.53 &1.61 & 1.59 & 1.55 & 1.58\\ 
2119 & c & 0.265& 0.003  &3.858& 0.63 &  0.61   &$-$ & $-$ & 0.59 & 1.61 & 1.62 &$-$ &  $-$ & 1.60\\ 
\tableline

\end{tabular}
\label{t:table3}
\normalsize

Notes: Puls. = {\it pulsational}, present paper; C98/P04 = Cassisi et al. (1998), and Pietrinferni et al. (2004); DC02/05= D'Antona 
et al. (2002), and Caloi \& D'Antona (2005); SC98 = Sweigart \& Catelan (1998), and Catelan et al. (1998); V00
= VandenBerg et al. (2000)\\
Masses and luminosities are in solar units. 
\vspace{0.5cm}

\end{table*}

\begin{table*}[ht]
\caption{Comparison with the intrinsic parameters derived 
from the Fourier decomposition of the light curve of the fundamental mode
pulsators}
\vspace{0.5cm}

\scriptsize
\begin{tabular}{rcccccccccc}
\tableline \tableline
Star&type& P   & $M/M_{\odot}$ & $M/M_{\odot}$ & $\log{L/L_{\odot}}$ & $\log{L/L_{\odot}}$ & $T_e$ & 
$T_e$    &$<M_V>$&$<M_V>$\\ 
    &    & day &               &  Fourier      &                     &      Fourier        &       &
Fourier   &     & Fourier \\ 
\tableline
5902& ab & 0.570& 0.75& 0.71 & 1.78 & 1.68 & 6700 & 6471 &0.347$\pm$ 0.060 & 0.531$\pm$0.027\\
2525& ab & 0.616& 0.75& 0.69 & 1.78 & 1.69 & 6570 & 6397 &0.336$\pm$ 0.062 & 0.513$\pm$0.027\\ 
2249& ab & 0.610& 0.70& 0.73 & 1.76 & 1.69 & 6625 & 6397 &0.392$\pm$ 0.065 & 0.520$\pm$0.027\\
9660& ab & 0.622& 0.68& 0.66 & 1.73 & 1.67 & 6500 & 6310 &0.467$\pm$ 0.063 & 0.580$\pm$0.026\\ 
7325& ab & 0.487& 0.62& 0.66 & 1.65 & 1.61 & 6800 & 6561 &0.640$\pm$ 0.061 & 0.684$\pm$0.028\\ 
\tableline

\end{tabular}
\label{t:table4}
\normalsize
\vspace{0.5cm}

Notes: Stars \#2249 and \#9660 have large Dm values: Dm$\leq$4.769 and Dm$\leq$3.746,
respectively (see Sect. 6 in DF04).
\end{table*}

\begin{table*}[ht]

\caption{Comparison of the reddening values}
\vspace{0.5cm}

\scriptsize
\begin{tabular}{rccccccc}
\tableline \tableline
Star&type&$E(B-V)$&$E(B-V)$&$E(B-V)$&$E(B-V)$&$E(B-V)$&$<E(B-V)>$\\ 
    &    &  mag        & mag        &  mag          & mag   & mag & mag\\
    &    & {\it Pulsational}  & C03(strip) & C03(Sturch)   & U99   & A04 & Adopted   \\
    &    &             &            &               & (1)   & (2) & (3)\\ 
\tableline
5902 & ab & 0.06$\pm$0.086&$0.086 \pm 0.017$   & 0.108   &  $-$  & $-$   &0.094$\pm 0.017$\\ 
2525 & ab & 0.13$\pm$0.101&$0.116 \pm 0.017$   & 0.169   &  $-$  & $-$   &0.138$\pm 0.037$\\ 
7620 & ab & 0.10$\pm$0.106&$0.086 \pm 0.017$   & 0.104   &  $-$  & $-$   &0.094$\pm 0.012$\\ 
7477 & ab & 0.10$\pm$0.096&$0.116 \pm 0.017$   & 0.130   &  $-$  & $-$   &0.121$\pm 0.011$\\ 
2249 & ab & 0.13$\pm$0.097&$0.086 \pm 0.017$   & 0.134   &  $-$  & $-$   &0.107$\pm 0.033$\\ 
9660 & ab & 0.14$\pm$0.097&$0.116 \pm 0.017$   & 0.100   &  $-$  & $-$   &0.110$\pm 0.012$\\ 
7325 & ab & 0.10$\pm$0.096&$0.116 \pm 0.017$   & 0.112   &  $-$  & $-$   &0.114$\pm 0.004$\\ 
2024 & ~c & 0.13$\pm$0.103&$0.116 \pm 0.017$   & $-$	 &  $-$  & $-$   &0.116$\pm 0.010$\\ 
4179 &~c  & 0.11$\pm$0.093&$0.086 \pm 0.017$   & $-$	 &  $-$  & $-$   &0.087$\pm 0.017$\\ 
8837 &~c  & 0.11$\pm$0.102&$0.116 \pm 0.017$   & $-$	 &  $-$  & 0.140 &0.116$\pm 0.004$\\ 
2623 &~c  & 0.07$\pm$0.104&$0.116 \pm 0.017$   & $-$	 &  $-$  & $-$   &0.115$\pm 0.033$\\ 
27697&~c  & 0.13$\pm$0.104&$0.116 \pm 0.017$   & $-$	 & 0.144$\pm 0.02$ & $-$  &0.116$\pm 0.010$\\ 
2517 & ~c & 0.14$\pm$0.100&$0.086 \pm 0.017$   & $-$	 &  $-$  & $-$   &0.088$\pm 0.038$\\
2119 & ~c & 0.10$\pm$0.113&$0.116 \pm 0.017$   & $-$	 & 0.144$\pm 0.02$ & $-$  &0.116$\pm 0.011$\\ 

\tableline
\end{tabular}
\label{t:table5}
\normalsize
\vspace{0.5cm}

Notes: (1) U99 reddening is the average value over the entire field OGLE-II LMC\_SC21, that includes 
about 40\% of C03 Field A; (2) A04 color excess is the average values from 330 RRc stars
spread over 16 MACHO fields close to the LMC bar; (3) weighted average of {\it pulsational},
C03(strip) and  C03(Sturch) reddenings.  
\vspace{0.5cm}
\end{table*}


\clearpage
\begin{references}
\reference{} Alcock, C., et al. 2004, \aj, 127, 334 (A04)
\reference{} Alves, D.R. 2004, New Astronomy Reviews, Volume  48, Issue 9, pages
  659-665 (astro-ph/0310673)
\reference{} Blazhko, S. 1907, Astron. Nachr., 175, 325 
\reference{} Bono, G., Castellani, V., \& Marconi, M. 2000, ApJ, 532, L129 (BCM00)
\reference{} Bono, G., Castellani, V., \& Marconi, M. 2002, ApJ, 565, L83 (BCM02) 
\reference{} Bono, G., Marconi, M., \& Stellingwerf 1999, ApJS, 122, 167
\reference{} Bono, G.,  Caputo, F., Cassisi, S., Castellani, V., Marconi, M., \&
Stellingwerf, R.F. 1997, ApJ, 477, 346
\reference{} Bono, G.,  Caputo, F., Castellani, V., Marconi, M., Storm, J., \& Degl'Innocenti, S. 2003, MNRAS, 344, 1097
\reference{} Bono, G., \& Stellingwerf, R.F. 1994, ApJS, 93, 233  
\reference{} Cacciari, C. 1999, in  ASP Conf. Ser. 167, Harmonizing Cosmic 
            Distance Scales in a Post-Hipparcos Era, ed. D. Egret, \& A. Heck
	    (San Francisco: ASP), 140
\reference{} Cacciari, C., Corwin, T.M., \& Carney, B.W. 2004, AJ, in press (astro-ph/0409567) 
\reference{} Caloi, V., D'Antona, F., \& Mazzitelli, I. 1997, A\&A, 320, 823
\reference{} Caloi, V., \& D'Antona, F. 2005, private communication 
\reference{} Caputo, F., De Rinaldis, A., Manteiga, M., Pulone, L., \& Quarta, M.L. 1993, A\&A, 276, 41
\reference{} Caputo, F., Castellani, V., Marconi, M., \& Ripepi, V. 2000, MNRAS, 316, 819 
\reference{} Cassisi, S., Castellani, V., Degl'Innocenti, S., \& Weiss, A. 1998,
   A\&AS, 129, 627 
\reference{} Cassisi, S., Castellani, M., Caputo, F., S., \& Castellani, V. 2004, 
   A\&A, 426, 641 
\reference{} Carretta, E., Gratton, R.G., Clementini, G., \& Fusi Pecci, F. 2000,
    \apj, 533, 215	        
\reference{} Castellani, V., Degl'Innocenti, S., \& Marconi, M. 2002, in ASP  Conf. Ser. 265, Omega Centauri, A Unique Window into Astrophysics, ed. F. van Leeuwen, J. D. Hughes, \& G. Piotto (San Francisco: ASP), 193
\reference{} Castelli, F., Gratton, R.G., \& Kurucz 1997a, A\&A, 318, 841    
\reference{} Castelli, F., Gratton, R.G., \& Kurucz 1997b, A\&A, 324, 432 
\reference{} Catelan, M., Borissova, J., Sweigart, A.V., \& Spassova, N. 1998,
            \apj, 494, 265
\reference{} Clementini, G., et al. 2000, \aj, 120, 2054
\reference{} Clementini, G., Gratton, R.G., Bragaglia, A., 
    Carretta, E., Di Fabrizio, L., \& Maio, M. 2003,\aj, 125, 1309 (C03) 
\reference{} Dambis, A.K. 2004, in EAS Publications Series, Galactic Dynamics, ed. C. Boily,
P. Patsis, C. Theis, S. Portegies Zwart, \& R. Spurzem, (EDP Science 2004), in press 
(astro-ph/0303463) 
\reference{} D'Antona, F., Caloi, V., Montalban, J., Ventura, P., \& Gratton, R.
 2002, A\&A, 395, 69    
\reference{} Di Criscienzo, M., Marconi, M., \& Caputo, F. 2004, ApJ, 612, 1092 
\reference{} Di Fabrizio, L. 1999, Laurea Degree Thesis, University of Bologna
\reference{} Di Fabrizio, L., et al.2002, MNRAS 336, 841 
\reference{} Di Fabrizio, L., Clementini, G., Maio, M., Bragaglia, A., 
    Carretta,E.,  Gratton, R.G., Montegriffo, P., \& Zoccali, E. 2004, A\&A, 
    in press (DF04, astro-ph/0409758)
\reference{} Freedman, W.L., et al. 2001, ApJ, 553, 47
\reference{} Fernley, J.A., Barnes, T.G., Skillen, I., Hawley, S.L., Hanley,
 C.J., Evans, D.W., Solano, E., \& Garrido, R. 1998a, A\&A, 330, 515   
\reference{} Fernley, J.A., Carney, B.W., Skillen, I., Cacciari, C., \& Janes, K. 1998b,
  MNRAS, 293, L61
\reference{} Gratton R.G., Bragaglia, A., Clementini, G., Carretta, E., 
    Di Fabrizio, L., Maio, M., \& Taribello, E. 2004, A\&A, 421, 937 (G04)
\reference{} Harris,W.E. 1996, \aj, 112, 1487 
\reference{} Jurcsik, J., \& Kov\`acs, G. 1996, A\&A, 312, 111
\reference{} Keller, S.C., \& Wood, P.R. 2002, ApJ, 578, 144 
\reference{} Lee, Y.W., Ddemarque, P., \& Zinn, R. 1990, ApJ, 350, 155
\reference{} Marconi, M., Caputo, F., Di Criscienzo, M., \& Castellani, M. 2003, ApJ, 596, 299
\reference{} Pietrinferni, A., Cassisi, S., Salaris, M., \& Castelli, F. 2004, ApJ,  612, 168
\reference{} Popowski, P. 2000, \apj, 528, L9
\reference{} Rastorguev, A.S., Dambis, A.K., \& Zabolotskikh, M.V. 2004, in 
The three Dimensional Universe with GAIA,  Paris-Meudon, in press
\reference{} Rey, S-C., Lee, Y-W., Joo, J-M., Walker, A., \& Baird, S. 2000,
 \aj, 119, 1824
\reference{} Rich, R.M., Corsi, C.E., Bellazzini, M., Federici, L., Cacciari, C.,
\& Fusi Pecci, F. 2001, in 
  Extragalactic Star Clusters, ed. E.K. Grebel, D. Geisler, \& D. Minniti
  (San Francisco:ASP, IAU Symp., 207, 140
\reference{} Rich, R.M., Corsi, C.E., Cacciari, C., Federici, L., 
 Fusi Pecci, F., Djorgovski, S.G., \& Freedman W. 2005, \aj, submitted
\reference{} Sandage, A. 1990, \apj, 350, 603
\reference{} Sandage, A. 1993, \aj, 106, 703
\reference{} Stellingwerf, R.F. 1982, ApJ, 262, 330
\reference{} Sturch, C. 1966, ApJ, 143, 774
\reference{} Sweigart, A.V., \& Catelan, M. 1998, \apj, 501, L63
\reference{} Tammann, G.A., Sandage, A., \& Reindl, B. 2003, A\&A, 404, 423 
\reference{} Udalski, A. 2000, \apj, 531, L25 
\reference{} Udalski, A., Szyma\'nski, M., Kubiak, M., 
 Pietrzy\'nski, G., Soszy\'nski, I., Wo\'zniak, P., \& Zebrun\'n, K. 1999, 
 Acta Astron., 49, 201 (U99)
\reference{} VandenBerg, D.A., Swenson, F.J., Rogers, F.J., Iglesias, C.A. \& Alexander, D.R. 2000, ApJ, 532, 430
\reference{} Walker, A.R. 1999, in Astrophysics and Space Science Library 237,
              Post-Hipparcos Cosmic Candles, ed. A. Heck, \& F. Caputo, 
	      (Kluwer Academic Publishers), 125  
\reference{} Walker, A.R. 2003, in Lecture Notes in Physics 635, 
 Stellar Candles for the Extragalactic Distance Scale, ed. 
    D. Alloin, \& W. Gieren, (Berlin:Springer), 265 
\reference{} Wood, P.R., Arnold, A., \& Sebo, K.M. 1997, ApJ, 485, L52 (WAS)
\reference{} Zaritsky, D. 1999, AJ, 118, 2824 
\reference{} Zaritsky, D., Harris, J., Thompson, I.B., \& Grebel, E.K. 2004, 
 AJ, in press (astro-ph/0407006)
\reference{} Zinn, R., \& West, M.J. 1984, \apjs, 55, 45
\end{references}
\end{document}